\documentclass[preprint,12pt]{elsarticle}

\usepackage{graphicx}

\usepackage{amssymb}

\usepackage{amsmath}
\usepackage{subfigure}
\usepackage{xspace}
\usepackage{cleveref}

\newcommand{\ie}{i.e.}

\newcommand{\given}{\! \mid \!}
\newcommand{\xbar}{\bar{x}}

\newcommand{\puncspace}{\enspace}
\newcommand{\mathcomma}{\puncspace ,}
\newcommand{\mathperiod}{\puncspace .}

\newcommand{\pa}[1]{\ensuremath{\mathrm{Pa}(#1)}} 

\newcommand{\set}[1]{\ensuremath{ \left \{#1 \right \}}}

\newcommand{\rhoest}{\hat{\rho}}

\newcommand{\meanval}[1]{\ensuremath{\xbar_{#1}}}

\newcommand{\maplinear}[2]{\ensuremath{A^{#1}_{#2}}}
\newcommand{\mapconst}[2]{\ensuremath{B^{#1}_{#2}}}
\newcommand{\mapfunc}[1]{\ensuremath{g\left(#1\right)}}
\newcommand{\act}[2]{\ensuremath{a^{#1}_{#2}}}

\newcommand{\decoder}[2]{\phi^{#1}_{#2}\left(x_{#1}\right)}
\newcommand{\decodeconst}[2]{\xbar^{#1}_{#2}}
\newcommand{\coupling}[2]{\ensuremath{K_{#1#2}}}

\newcommand{\externsignal}[2]{\ensuremath{h^{#1}_{#2}}}
\newcommand{\stablewts}[3]{\ensuremath{S^{#1}_{#2#3}}}
\newcommand{\selfwts}[3]{\ensuremath{T^{#1}_{#2#3}}}

\newcommand{\ffwts}[4]{\ensuremath{U^{#1#2}_{#3#4}}}
\newcommand{\fbwts}[4]{\ensuremath{V^{#2#1}_{#3#4}}}
\newcommand{\indirectwts}[5]{\ensuremath{W^{#1#2#3}_{#4#5}}}

\newcommand{\wtdefwrap}[4]{\maplinear{#1}{#2}{#4}\decodeconst{#1}{#3}}
\newcommand{\varinv}[1]{\frac{1}{\sigma_{#1}^2}}
\newcommand{\couplevar}[2]{\varinv{#1}\coupling{#1}{#2}}

\crefname{equation}{Eq.}{Eqs.}
\crefname{figure}{Fig.}{Figs.}
\crefname{subfigure}{Fig.}{Figs.}
\crefname{section}{Section}{Sections}
\crefname{subsection}{Section}{Sections}
\crefname{subsubsection}{Section}{Sections}
\crefname{table}{Table}{Tables}

\graphicspath{{pdffigs/}{epsfigs/}{mpsfigs/}}

\journal{arXiv}

\begin{document}

\begin{frontmatter}



\title{Designing neural networks that process mean values of random variables}


\author[ait]{Michael J. Barber \corref{cor1}}
\ead{michael.barber@ait.ac.at}
\author[washu]{John W. Clark}

\address[ait]{AIT Austrian Institute of Technology GmbH, Foresight and Policy Development Department, Vienna, Austria}
\address[washu]{Department of Physics,
    Washington University,
    Saint Louis, MO 63130}
    
\cortext[cor1]{Corresponding author}

\begin{abstract}
    We introduce a class of neural networks derived from probabilistic
    models in the form of Bayesian networks.  By imposing additional
    assumptions about the nature of the probabilistic models
    represented in the networks, we derive neural networks with
    standard dynamics that require no training to determine the
    synaptic weights, that perform accurate calculation of the mean 
    values of the random variables, that can pool multiple sources of evidence,
    and that deal cleanly and consistently with inconsistent or
    contradictory evidence.  The presented neural networks capture
    many properties of Bayesian networks, providing distributed
    versions of probabilistic models.
\end{abstract}


\end{frontmatter}


\section{Introduction}\label{sec:intro}

Artificial neural networks are noted for their ability to learn functional relationships from observed data. 
Unfortunately, a trained neural network is typically a black box, so that it can be quite difficult to determine what function is actually represented by the network. 
In numerous cases, neural networks have been related to probabilistic models, with either the trained network retrospectively given a probabilitic interpretation or the training process itself explicitly based on a probabilistic strategy.
Alternatively, a constructive approach can be taken to exploring representation of probabilistic models in neural networks, encoding pre-specified probabilistic models into network weights. 
A key challenge in this alternate approach is to produce reasonable neural networks, allowing a suitably broad class of probabilistic models to be encoded into neural networks with recognizable architectures and dynamics. 
Towards this end, in this paper we formulate and characterize an encoding method that handles a restricted class of probabilistic models and allows calculation, without training, of neural networks that accurately process the mean values of the random variables with the usual neural activation of a weighted sum of the 
neural inputs transformed with a nonlinear activation function.

Probabilistic formulations of neural information processing have been
explored along a number of avenues.  
One of the earliest such analyses 
showed that the original Hopfield neural network implements, in 
effect, Bayesian inference on analog quantities in terms of probability densities 
\citep{anderson/abrahams:1987}.   
\citet{zemel/etal:1998} have
investigated population coding of probability distributions, but with 
different representations and dynamics
than those we consider in this paper.  Several extensions of this representation 
scheme have been developed 
\citep{zemel:1999,zemel/dayan:1999,yang/zemel:2000} that 
feature information propagation between interacting neural populations.  
Additionally, several ``stochastic 
machines'' \citep{haykin:1999} have been formulated, including Boltzmann 
machines \citep{hinton/sejnowski:1986}, sigmoid belief networks 
\citep{neal:1992}, and Helmholtz machines \citep{dayan/hinton:1996}.  Stochastic 
machines are built of stochastic neurons that occupy  one of two 
possible states in a probabilistic manner. Learning rules for 
stochastic machines enable such systems to model the underlying probability 
distribution of a given data set.

Additionally, the connection between neural networks and probabilistic
models represented specifically as Bayesian networks
\citep{pearl:1988,smyth/etal:1997} has been explored along two main
lines. In one approach, the neural network architecture and activation
dynamics are specified, with a learning rule used that attempts to
capture the appropriate Bayesian network in the synaptic weights
based on observed patterns \citep{neal:1992,George2005Hierarchical}.
In a second approach, a prespecified Bayesian network is transformed
into a neural network using an encoding process
\citep{BarClaAnd:2003,BarClaAnd:2003a}. While specific Bayesian
networks are readily given in the latter approach, the neural
architecture and dynamics arise from the encoding and need not match
existing definitions. In particular, instead of the usual weighted
sum of neural activation values passed through a nonlinear activation
function, the encoding process can produce neural networks depending
on multiplicative interactions between neural activities.

In this work, we further explore the latter, encoding-based approach. By imposing 
additional, strong assumptions about the originating Bayesian network, we develop 
neural networks processing mean values of analog 
variables, where all weights are calculated and no learning process is needed. 
The random variables are assumed to be normally distributed, which results in only the 
mean values being accurately encoded into the neural network. 
The resulting 
dynamics are of the usual form for neural networks, \ie, a weighted sum of the 
neural inputs transformed with a nonlinear activation function. 

We 
begin with a brief summary of the key relevant properties of Bayesian 
networks in \cref{sec:bbns}. We describe a procedure for generating and 
evaluating the neural networks in \cref{sec:mvbns}, and apply the 
procedure to several examples in \cref{sec:applications}. 

\section{Bayesian Networks}\label{sec:bbns}

\begin{figure}[tbp]
\centering
\includegraphics[scale=.5]{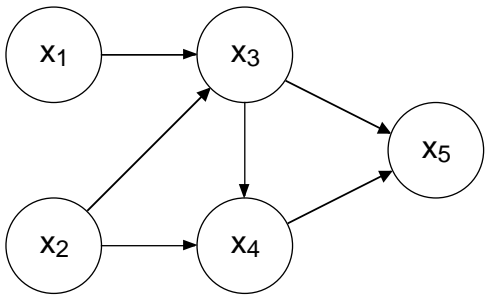}
\caption{A Bayesian network.  Evidence about any of the
random variables influences the likelihood of the 
remaining random variables. 
In a straightforward terminology, the node at the tail 
of an arrow is a parent of the child node at the head of the arrow, e.g.
\(X_4\) is a parent of \(X_5\) and a child of both \(X_2\) and
\(X_3\).
From the structure of the graph, we can see the conditional independence 
relations in the probabilistic model.  For example, \(X_5\) is 
independent of \(X_1\) and \(X_2\) given \(X_3\) and \(X_4\).  
}
\label{fig:bbn}
\end{figure}

Bayesian networks \citep{pearl:1988,smyth/etal:1997} are directed 
acyclic 
graphs that represent 
probabilistic models (\cref{fig:bbn}).  
Each node represents 
a random variable, and the arcs signify the presence of dependence between the linked variables.  The strengths of these 
influences are defined using conditional probabilities.  We additionally take the 
direction of a particular link to indicate the direction of causality (or, 
more simply, relevance), with an arc pointing from cause to effect; in this form, the 
Bayesian network is also called a causal network.

Multiple sources of evidence about the random variables are conveniently
handled using Bayesian networks. The belief, or degree of confidence, in particular 
values of the random
variables is determined as the likelihood of the value given evidentiary
support provided to the network.
There are two types of support that arise from the
evidence: predictive support, which propagates from cause to effect 
along the direction of the arc, and retrospective support, which
propagates from effect to cause, opposite to the direction of the
arc. 

Bayesian networks have two properties that we will find very useful, 
both of which stem 
from the dependence relations shown by the graph structure.  First, the value 
of a node $X$ is not dependent upon all of the other graph nodes.  
Rather, it depends only on a subset of the nodes, called a Markov 
blanket of $X$, that separates node $X$ from all the other nodes in 
the graph.  The Markov blanket of interest to us is readily determined from 
the graph structure.  It is comprised of the union of the direct parents 
of $X$, the 
direct successors of $X$, and all direct parents of the direct 
successors of $X$.  Second, the joint probability over the random 
variables is decomposable as
\begin{equation}
    P(x_{1},x_{2},\ldots,x_{n}) = \prod_{\mu=1}^{n} P(x_{\mu}\given 
    \pa{x_{\mu}}) \mathcomma
    \label{eq:bbndecompose}
\end{equation}
where \pa{x_{\mu}} denotes the (possibly empty) set of direct-parent nodes 
of $X_{\mu}$.  
This decomposition comes about from repeated application of Bayes' 
rule and from the structure of the graph.  

\section{Neural Network Model}\label{sec:mvbns}

We will develop  neural networks from
the set of marginal distributions \set{\rho(x_{\mu};t)}
so as to best match a desired probabilistic model \( \rho(x_1, x_2,
\ldots, x_D) \) over the set of random variables, which are
organized as a Bayesian network.  One or more of the variables \( x_{\mu} \)  must be
specified as evidence in the Bayesian network. To facilitate the development of general
update rules, we do not distinguish between evidence and 
non-evidence nodes in our notation.  

Our general approach will be to minimize the difference between 
a probabilistic model  \( \rho(x_1, x_2, \ldots, x_D) \) and an estimate 
of the probabilistic model \( \rhoest(x_1, x_2, \ldots, x_D) \). For 
the estimate, we utilize 
\begin{equation}
    \rhoest(x_1, x_2, \ldots, x_D) = 
    \prod_{\alpha} \rho(x_{\alpha};t) 
    \mathperiod
    \label{eq:estnaive}
\end{equation}
This is a  so-called naive estimate, wherein the random variables are 
assumed to be independent. We will place further constraints on the 
probabilistic model and representation to produce neural networks with 
the desired dynamics.  

\begin{figure}[tbp]
\centering
\includegraphics[width=\columnwidth]{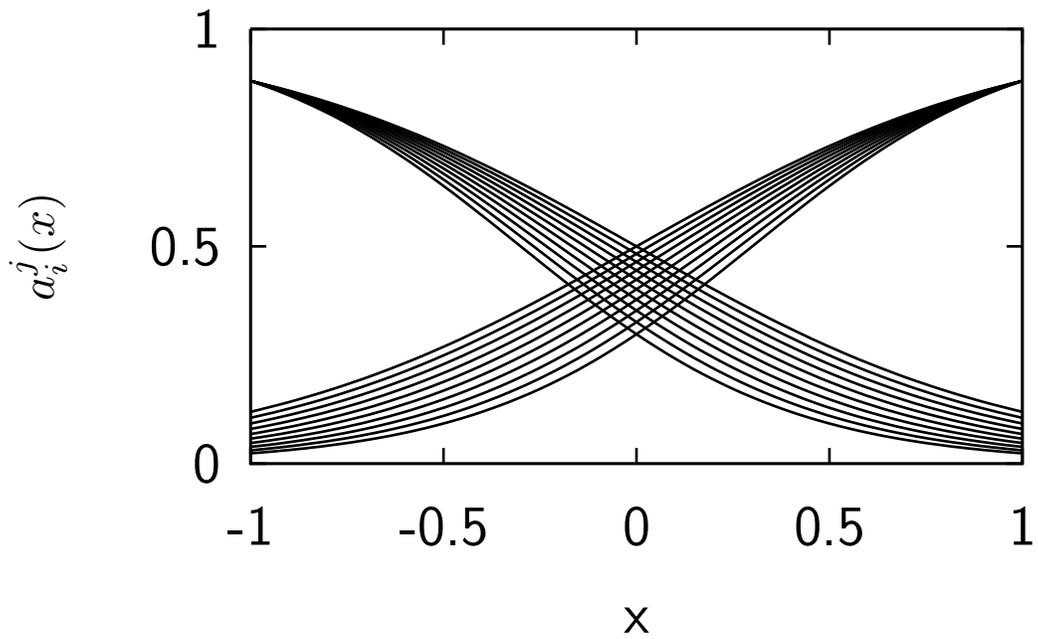}
\caption{The mean values of the random variables are encoded into
the firing rates of populations of neurons.  A population of twenty
neurons with sigmoidal responses is associated with each 
random variable.  The neuronal responses \act{\alpha}{i} are fully determined by 
a
single input \( \xi \), which we interpret as the mean value of a
density.  
The form of the neuronal transfer functions can be altered without
affecting the general result presented in this work.
}
\label{fig:firingrates}
\end{figure}

The first assumption we make is that the populations of neurons only
need to accurately encode the mean values of the random variables,
rather than the complete densities.  We take the firing rates of the
neurons representing a given random variable \(X_{\alpha}\) to be 
functions of the mean value \meanval{\alpha}(t)
\begin{equation}
    \act{\alpha}{i}(t) = \mapfunc{\maplinear{\alpha}{i}
    \meanval{\alpha}(t) + \mapconst{\alpha}{i}}
    \mathcomma 
    \label{eq:neurresponses}
\end{equation}
where 
\maplinear{\alpha}{i}
and 
\mapconst{\alpha}{i} 
are parameters describing 
the response properties of neuron \( i \) of the population
representing random variable \( X_{\alpha} \).  The activation function \(g\)
is in general nonlinear; in this work, we take \(g\) to be the logistic
function,
\begin{equation}
    \mapfunc{x} = \frac{1}{1+\exp\left(-x\right)}
    \mathperiod
\end{equation}
We use a set of neural response function (\cref{fig:firingrates}) similar to ones
from work on population-temporal coding that supported manipulation of mean 
values \citep{EliAnd:1999,EliAnd:2002}. 
We can make use 
of \cref{eq:neurresponses} to directly encode mean values into 
neural activation states, providing a means to specify the value of the
evidence nodes in the NBN. 

Using \cref{eq:neurresponses}, we  derive an update rule
describing the neuronal dynamics,
obtaining (to first order in \(\tau\))
\begin{equation}
    \act{\alpha}{i}(t +\tau) = 
    \mapfunc{\maplinear{\alpha}{i} \meanval{\alpha}(t)
    +\tau \maplinear{\alpha}{i}  \frac{d\meanval{\alpha}(t)}{dt}  +
    \mapconst{\alpha}{i} }
    \mathperiod
    \label{eq:updaterule}
\end{equation}
Thus, if we can determine how \meanval{\alpha} changes with time, we can
directly determine how the neural activation states change with time.

The mean value \(\meanval{\alpha}(t)\) can be determined from the firing
rates as the expectation value of the random variable \(X_{\alpha}\) with
respect to a density \(\rho(x_{\alpha}; t)\) represented in terms of some
decoding functions 
\set{\decoder{\alpha}{i}}
The density is recovered using
the relation
\begin{equation}
	\rho(x_{\alpha};t) = \sum_i \act{\alpha}{i} \decoder{\alpha}{i}
    \mathperiod
    \label{eq:decodingrule}
\end{equation}
The decoding functions are constructed so as to minimize the difference
between the assumed and reconstructed densities (discussed in detail in 
\cite{barber/clark/anderson:2003a}).

With representations as given in 
\cref{eq:decodingrule}, we have
\begin{align}
    \meanval{\alpha}(t) & = \int x_\alpha \rho\left(x_{\alpha}; t\right)
    \, dx_{\alpha} \nonumber \\
    & = \sum_i \act{\alpha}{i}(t) \decodeconst{\alpha}{i} 
    \mathcomma 
	\label{eq:expectvalues}
\end{align}
where we have defined
\begin{equation}
    \decodeconst{\alpha}{i} = \int x_{\alpha} \decoder{\alpha}{i}\, dx_{\mu} 
    \mathperiod
\end{equation}
Although we used the decoding functions \(\decoder{\alpha}{i}\) to 
calculate the 
parameters \(\decodeconst{\alpha}{i}\), they can in practice be found directly 
so that the
relations in \cref{eq:neurresponses,eq:expectvalues} are mutually 
consistent.

We take the densities \(\rho(x_{\alpha}; t)\) to be normally distributed
with the form
\( \rho(x_{\alpha};t) \equiv \rho(x_{\alpha}; \meanval{\alpha}(t)) 
= N(x_{\alpha}; \meanval{\alpha}(t),
\sigma_{x_{\alpha}}^2)\).  Intuitively, we might expect that the variance
\( \sigma_{x_{\alpha}}^2\) should be small so that the mean value is coded
precisely, but we will see that the variances have no significance in
the resulting neural networks.

The second assumption we make 
is that interactions between the
nodes are linear:
\begin{equation}
    x_{\beta} = \sum_{\alpha}
	\coupling{\beta}{\alpha}
	x_{\alpha} 
    \mathperiod
\end{equation}
Utilizing the causality relations given by the Bayesian network,
we require that  \(\coupling{\beta}{\alpha} \neq 0\) only if \(X_{\beta}\) is 
a child node of \(X_{\alpha}\) in the network graph.
To represent the linear interactions as a probabilistic model, we
take the normal distributions \(\rho(x_{\beta} \given \pa{x_{\beta}}) = 
N(x_{\beta};  \sum_{\alpha} 
\coupling{\beta}{\alpha}x_{\alpha}, \sigma_{\beta}^2 ) \) for the
conditional probabilities.

For nodes in the Bayesian network which have no parents, the conditional
probability \(\rho(x_{\beta} \given \pa{x_{\beta}})\) is just the
prior probability distribution \(\rho(x_{\beta})\). We utilize the 
same rule to define the prior probabilities  as to define 
the conditional probabilities. For parentless nodes,
the prior is thus normally distributed with 
zero mean, \(\rho(x_{\beta}) = N(x_{\beta}; 0 , \sigma_{\beta}^2 )\).

We use the relative entropy \citep{papoulis:1991} as a measure of the
``distance'' between the joint distribution
describing the probabilistic model \( \rho(x_1, x_2,
\ldots, x_D)\) and the 
density estimated from the neural 
network \(\rhoest(x_1, x_2, \ldots, x_D)\).
Thus, we minimize
\begin{equation}
    E = - \int \rhoest(x_1, x_2, \ldots,x_D) \log \left (
	\frac
	{\rho(x_1, x_2, \ldots,x_D)} 
    {\rhoest(x_1, x_2, \ldots,x_D)}
    \right ) \,
    dx_1 dx_2\cdots dx_D
\end{equation}
with respect to the mean values \meanval{\alpha}.  By making use of the
gradient descent prescription 
\begin{equation}
    \frac{d\meanval{\gamma}}{dt} = -\eta \frac{\partial E}{\partial \meanval{
\gamma}}
\end{equation}
and the decomposition property for Bayesian networks given by \cref{eq:bbndecompose},
we obtain the update rule for the mean
values,
\begin{equation}
    \frac{d\meanval{\gamma}}{dt} = \frac{\eta}{\sigma_{\gamma}^2} 
    \left( 
    \sum_{\beta}
    \coupling{\gamma}{\beta}\meanval{\beta} - \meanval{\gamma}\right) 
    -\eta \sum_{\beta}
    \frac{\coupling{\beta}{\gamma}}{\sigma_{\beta}^2}
    \left ( \sum_{\alpha} 
    \coupling{\beta}{\alpha}\meanval{\alpha} - \meanval{\beta} \right )
    \mathperiod
    \label{eq:finalupdate}
\end{equation}
Because the coupling parameters \coupling{\alpha}{\beta} are nonzero 
only when \(X_{\alpha}\) is a parent of \(X_{\beta}\), generally only a 
subset of the mean values contributes to updating \(\xbar_{\gamma}\)
in \cref{eq:finalupdate}. In terms of the 
Bayesian network graph structure, the only contributing values 
come from the parents 
of \(X_{\gamma}\), the children of  \(X_{\gamma}\), and the parents 
of the children of \(X_{\gamma}\); this is identical to the Markov blanket
discussed in \cref{sec:bbns}.

The update rule for the neural activities is obtained by combining
\cref{eq:updaterule,eq:expectvalues,eq:finalupdate}, resulting in
\begin{equation}
    \act{\gamma}{i}(t+\tau) = \mapfunc{ 
        \sum_j \stablewts{\gamma}{i}{j} \act{\gamma}{j}(t) +
        \mapconst{\gamma}{i}+ \eta \tau \externsignal{\gamma}{i}(t)}
		\mathperiod
		\label{eq:neuralupdate}
\end{equation}
The quantity \(\sum_j \stablewts{\gamma}{i}{j} \act{\gamma}{j}(t) +
\mapconst{\gamma}{i}\) serves to stabilize the activities of the neurons 
representing \(\rho(x_{\gamma})\),
while  
\begin{equation}
    \externsignal{\gamma}{i} (t)= 
    \sum_j \selfwts{\gamma}{i}{j} \act{\gamma}{j}(t) +
    \sum_\beta \sum_j \left(\ffwts{\gamma}{\beta}{i}{j} +
    \fbwts{\gamma}{\beta}{i}{j}\right)\act{\beta}{j}(t) +
    \sum_{\alpha,\beta} \sum_j 
    \indirectwts{\gamma}{\beta}{\alpha}{j}{i} \act{\alpha}{j}(t) 
    \label{eq:neuralinput}
\end{equation}
drives changes in \(\act{\gamma}{i}(t)\)
based on the densities represented by other nodes of the Bayesian network.
The synaptic weights of the neural network are
\begin{align}
    \stablewts{\gamma}{i}{j} & = 
        \wtdefwrap{\gamma}{i}{j}{} \mathcomma\\
    \selfwts{\gamma}{i}{j} & = 
        -\wtdefwrap{\gamma}{i}{j}{\varinv{\gamma}} \mathcomma\\
    \ffwts{\gamma}{\beta}{i}{j} & = 
        \wtdefwrap{\gamma}{i}{j}{\couplevar{\gamma}{\beta}}  \mathcomma\\
    \fbwts{\gamma}{\beta}{i}{j} & = 
        -\wtdefwrap{\gamma}{i}{j}{\couplevar{\beta}{\gamma}} \mathcomma\\
    \indirectwts{\gamma}{\beta}{\alpha}{j}{i} & = 
        \wtdefwrap{\gamma}{i}{j}
		{\couplevar{\gamma}{\beta}\coupling{\beta}{\alpha}}
	\label{eq:indirectwts}
\end{align}

The foregoing provides an algorithm for generating and evaluating 
neural networks that process mean values of random variables.  To 
summarize,
\begin{enumerate}
    \item  Establish independence relations between model variables.  
    This may be accomplished by  using a graph to organize the 
    variables.
    \item  Specify the \coupling{\alpha}{\beta} to quantify the relations 
between 
    the variables.
    \item  Assign network inputs by encoding desired values into 
    neural activities using \cref{eq:neurresponses}.
    \item  Update other neural activities
    using the update rule in \cref{eq:neuralupdate} and the supporting definitions 
    in \crefrange{eq:neuralinput}{eq:indirectwts}. 
    \item  Extract the expectation values of the variables from the neural 
    activities using \cref{eq:expectvalues}.
\end{enumerate}

\section{Examples}\label{sec:applications}

As a first example, we apply the algorithm to the Bayesian network shown in 
\cref{fig:bbn}, with firing rate profiles as shown in
\cref{fig:firingrates}. Specifying \(x_1 = 1/2\) and \(x_2 =
-1/2\) as evidence, we find an excellent match between the mean values
calculated by the neural network and the directly calculated values
for the remaining nodes (\cref{table:comparison}).

\begin{table}[b]
    \centering
    \caption{The mean values decoded from the neural network closely
    match the values directly calculated from the linear relations.  The
    coefficients for the linear combinations were randomly selected,
    with values \( \coupling{3}{1} = -0.2163\), \(\coupling{3}{2} = -0.8328\), 
    \(\coupling{4}{2} =
    0.0627\), \(\coupling{4}{3} = 0.1438\), \(\coupling{5}{3} = -0.5732\), 
    and \(\coupling{5}{4} =
    0.5955\). } \label{table:comparison}
    \begin{tabular}{ccc}
    \hline\hline
	Node & Direct Calculation & Neural Network \\ \hline
	\(X_1\) &\ 0.5000 &\ 0.5000 \\
	\(X_2\) & -0.5000 & -0.5000\\
	\(X_3\) & \ 0.3083 & \ 0.3084\\
	\(X_4\) & \ 0.0130 & \ 0.0128\\
	\(X_5\) & -0.1690 & -0.1689 \\
	\hline \hline
    \end{tabular}
\end{table}

\begin{figure}[tbp]
    \centering
    \subfigure[]{
    	\includegraphics[scale=.5]{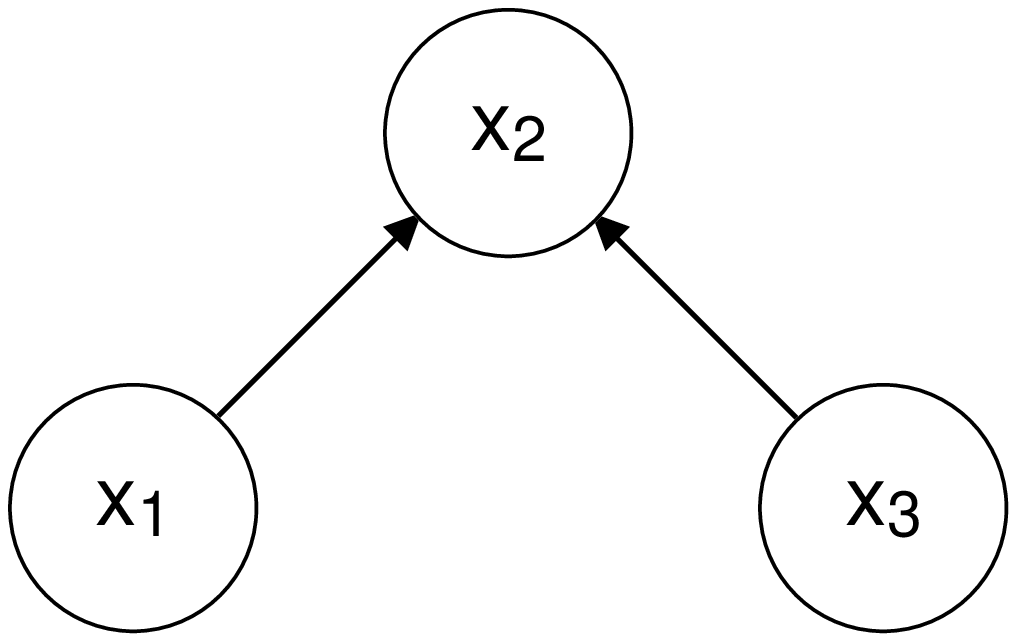}
    	\label{fig:tree1}
    }
    \hfil
    \subfigure[]{
    	\includegraphics[scale=.5]{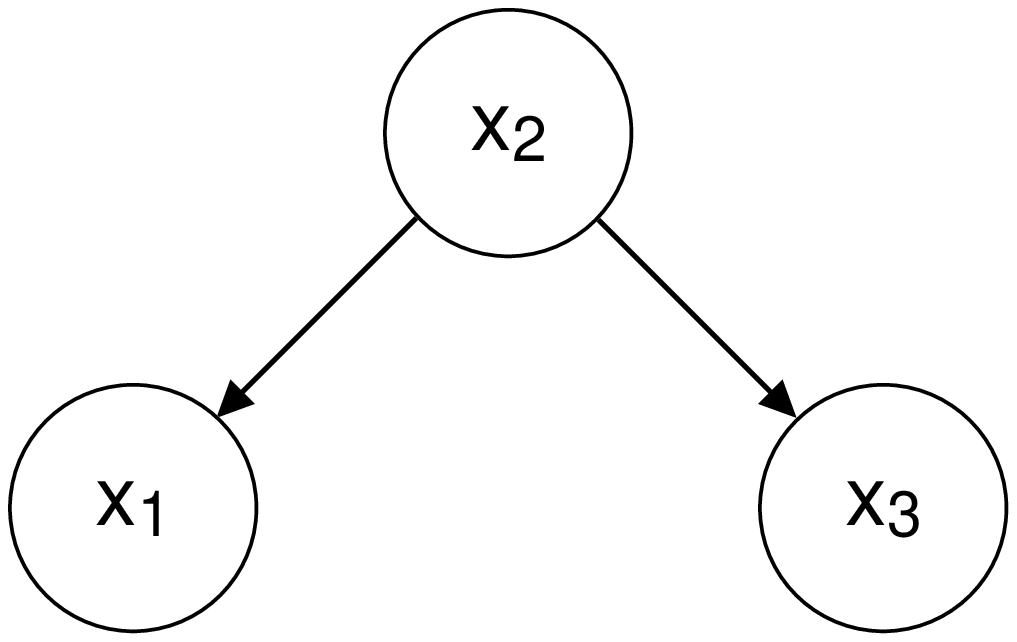}
    	\label{fig:tree2}
    }
    \caption{Simpler Bayesian networks.  Although the underlying undirected graph 
    structure is identical for these two networks, the direction of 
    the causality relationships between the variables are reversed.   
    The neural networks arising from the Bayesian networks thus have different 
    properties.}
    \label{fig:trees}
\end{figure}

We next focus on some simpler Bayesian networks to highlight certain 
properties of the resulting neural networks (which will again utilize 
the firing rate profiles shown in 
\cref{fig:firingrates}).  In 
\cref{fig:trees}, we present two Bayesian networks that relate three random 
variables in different ways.  The connection strengths are all taken 
to be unity in each graph, so that \( \coupling{2}{1} = \coupling{2}{3} 
= \coupling{1}{2} = 
\coupling{1}{3} = 1\).  

With the connection strengths so chosen, the two Bayesian networks have  
straightforward interpretations.  For the graph shown in 
\cref{fig:tree1}, \( X_{2} \) represents the sum of \( X_{1} \) 
and \( X_{3} \), while, for the graph shown in 
\cref{fig:tree2}, \( X_{2} \) provides a value which is 
duplicated in \( X_{1} \) and \( X_{3} \).  The different graph 
structures yield different neural networks; in particular, nodes 
\( X_{1} \) and \( X_{3} \) have direct synaptic connections in the 
neural network based 
on the graph in 
\cref{fig:tree1}, but no such direct weights exist in a second 
network based on
\cref{fig:tree2}.  Thus, specifying \( x_{1} = -1/4\) and 
\( x_{2} = 1/4 \) for the first network produces the expected result 
\( \xbar_{3} = -0.5000 \), but specifying \( x_{2} = 1/4 \) in the second 
network produces  
\( \xbar_{3} = 0.2500 \) regardless of the value (if any) 
assigned to \( x_{1} \).   

To further illustrate the neural network properties, we use
the graph shown in \cref{fig:tree2} to process 
inconsistent evidence.  Nodes \( X_{1} \) and \( X_{3} \) should 
copy the value in node \( X_{2} \), but we can specify any values we 
like as network inputs.  For example, when we assign \( x_{1} = -1/4 \) and 
\( x_{3} = 1/2 \), the neural network yields \( \xbar_{2} = 0.1250 \) for 
the remaining value.  
This is a typical and reasonable result, matching
the least-squares solution to the inconsistent problem.

\section{Conclusion}\label{sec:concl}

We have introduced a class of neural networks that
consistently mix multiple sources of evidence. The networks are
based on probabilistic models, represented in the graphical form of
Bayesian networks, and function based on traditional
neural network dynamics (\ie, a weighted sum of neural 
activation values passed through a nonlinear activation function).
We constructed the networks by
restricting the represented probabilistic models by introducing 
two auxiliary assumptions.  

First, we assumed that only the mean values of the
random variables need to be accurately represented, with higher
order moments of the distribution being unimportant. We introduced
neural representations of relevant probability density functions
consistent with this assumption.  Second, we assumed that the random
variables of the probabilistic model are linearly related to one another, 
and chose appropriate
conditional probabilities to implement these linear relationships. 

Using the representations suggested by our auxiliary assumptions, we
derived a set of update rules by minimizing the relative entropy
of an assumed density with respect to the density decoded from the neural
network.  In a straightforward fashion, the optimization procedure
yields neural weights and  dynamics that implement specified probabilistic 
relations, without the need for a training process.

The neural networks investigated in this work
captures many of the
properties of both Bayesian networks and traditional neural network models.  
In particular, 
multiple sources of evidence are consistently pooled based on local
update rules, providing a distributed version of a
probabilistic model. 

\section*{Acknowledments}
    This work was supported by the U.S. National Science 
    Foundation, by the Portuguese 
    Funda{\c c}\~ao para a Ci\^encia e a Technologia (FCT) under Bolsa
    SFRH/BPD/9417/2002, and by the Austrian Science Fund (FWF) under project P21450. 
    JWC also acknowledges support received from 
    the Funda{\c c}\~ao Luso-Americana para o Desenvolvimento (FLAD)
    and from the FCT for his participation in Madeira Math Encounters XXIII at
    the University of Madiera, where portions of the work were conducted.





\bibliographystyle{model1-num-names}
\bibliography{neurcomp,mjbpubs,references}

\begin{thebibliography}{18}
\expandafter\ifx\csname natexlab\endcsname\relax\def\natexlab#1{#1}\fi
\providecommand{\bibinfo}[2]{#2}
\ifx\xfnm\relax \def\xfnm[#1]{\unskip,\space#1}\fi
\bibitem[{Anderson and Abrahams(1987)}]{anderson/abrahams:1987}
\bibinfo{author}{C.~H. Anderson}, \bibinfo{author}{E.~Abrahams},
\newblock \bibinfo{title}{The {Bayes} connection},
\newblock in: \bibinfo{booktitle}{Proceedings of the {IEEE} First International
  Conference on Neural Networks}, volume \bibinfo{volume}{III},
  \bibinfo{organization}{IEEE}, \bibinfo{publisher}{{SOS} Print},
  \bibinfo{address}{San Diego, CA}, \bibinfo{year}{1987}, pp.
  \bibinfo{pages}{105--112}.
\bibitem[{Zemel et~al.(1998)Zemel, Dayan, and Pouget}]{zemel/etal:1998}
\bibinfo{author}{R.~S. Zemel}, \bibinfo{author}{P.~Dayan},
  \bibinfo{author}{A.~Pouget},
\newblock \bibinfo{title}{Probabilistic interpretation of population codes},
\newblock \bibinfo{journal}{Neural Comput.} \bibinfo{volume}{10}
  (\bibinfo{year}{1998}) \bibinfo{pages}{403--30}.
\bibitem[{Zemel(1999)}]{zemel:1999}
\bibinfo{author}{R.~S. Zemel},
\newblock \bibinfo{title}{Cortical belief networks},
\newblock in: \bibinfo{editor}{R.~Hecht-Neilsen} (Ed.),
  \bibinfo{booktitle}{Theories of Cortical Processing},
  \bibinfo{publisher}{Springer-Verlag}, \bibinfo{address}{Berlin},
  \bibinfo{year}{1999}.
\bibitem[{Zemel and Dayan(1999)}]{zemel/dayan:1999}
\bibinfo{author}{R.~S. Zemel}, \bibinfo{author}{P.~Dayan},
\newblock \bibinfo{title}{Distributional population codes and multiple motion
  models},
\newblock in: \bibinfo{editor}{M.~S. Kearns}, \bibinfo{editor}{S.~A. Solla},
  \bibinfo{editor}{D.~A. Cohn} (Eds.), \bibinfo{booktitle}{Advances in Neural
  Information Processing Systems 11}, \bibinfo{publisher}{MIT Press},
  \bibinfo{address}{Cambridge, MA}, \bibinfo{year}{1999}.
\bibitem[{Yang and Zemel(2000)}]{yang/zemel:2000}
\bibinfo{author}{Z.~Yang}, \bibinfo{author}{R.~S. Zemel},
\newblock \bibinfo{title}{Managing uncertainty in cue combination},
\newblock in: \bibinfo{editor}{S.~A. Solla}, \bibinfo{editor}{K.~R. Leen, T. K.
  and.~M{\"u}ller} (Eds.), \bibinfo{booktitle}{Advances in Neural Information
  Processing Systems 12}, \bibinfo{publisher}{MIT Press},
  \bibinfo{address}{Cambridge, MA}, \bibinfo{year}{2000}.
\bibitem[{Haykin(1999)}]{haykin:1999}
\bibinfo{author}{S.~Haykin}, \bibinfo{title}{Neural Networks: A Comprehensive
  Foundation}, \bibinfo{publisher}{Prentice Hall}, \bibinfo{address}{Upper
  Saddle River, NJ}, \bibinfo{edition}{second} edition, \bibinfo{year}{1999}.
\bibitem[{Hinton and Sejnowski(1986)}]{hinton/sejnowski:1986}
\bibinfo{author}{G.~E. Hinton}, \bibinfo{author}{T.~J. Sejnowski},
\newblock \bibinfo{title}{Learning and relearning in {Boltzmann} machines},
\newblock in: \bibinfo{editor}{D.~E. Rumelhart}, \bibinfo{editor}{J.~L.
  McClelland}, et~al. (Eds.), \bibinfo{booktitle}{Parallel Distributed
  Processing: Explorations in Microstructure of Cognition},
  \bibinfo{publisher}{MIT Press}, \bibinfo{address}{Cambridge, MA},
  \bibinfo{year}{1986}, pp. \bibinfo{pages}{282--317}.
\bibitem[{Neal(1992)}]{neal:1992}
\bibinfo{author}{R.~M. Neal},
\newblock \bibinfo{title}{Connectionist learning of belief networks},
\newblock \bibinfo{journal}{Artificial Intelligence} \bibinfo{volume}{56}
  (\bibinfo{year}{1992}) \bibinfo{pages}{71--113}.
\bibitem[{Dayan and Hinton(1996)}]{dayan/hinton:1996}
\bibinfo{author}{P.~Dayan}, \bibinfo{author}{G.~E. Hinton},
\newblock \bibinfo{title}{Varieties of {Helmholtz} machines},
\newblock \bibinfo{journal}{Neural Networks} \bibinfo{volume}{26}
  (\bibinfo{year}{1996}) \bibinfo{pages}{1385--1403}.
\bibitem[{Pearl(1988)}]{pearl:1988}
\bibinfo{author}{J.~Pearl}, \bibinfo{title}{Probabilistic Reasoning in
  Intelligent Systems: Networks of Plausible Inference},
  \bibinfo{publisher}{Morgan Kaufmann}, \bibinfo{address}{San Mateo, CA},
  \bibinfo{year}{1988}.
\bibitem[{Smyth et~al.(1997)Smyth, Heckerman, and Jordan}]{smyth/etal:1997}
\bibinfo{author}{P.~Smyth}, \bibinfo{author}{D.~Heckerman},
  \bibinfo{author}{M.~I. Jordan},
\newblock \bibinfo{title}{Probabilistic independence networks for hidden
  {Markov} probability models},
\newblock \bibinfo{journal}{Neural Comput.} \bibinfo{volume}{9}
  (\bibinfo{year}{1997}) \bibinfo{pages}{227--69}.
\bibitem[{George and Hawkins(2005)}]{George2005Hierarchical}
\bibinfo{author}{D.~George}, \bibinfo{author}{J.~Hawkins},
\newblock \bibinfo{title}{A hierarchical bayesian model of invariant pattern
  recognition in the visual cortex},
\newblock volume~\bibinfo{volume}{3}, pp. \bibinfo{pages}{1812--1817}.
\bibitem[{Barber et~al.(2003{\natexlab{a}})Barber, Clark, and
  Anderson}]{BarClaAnd:2003}
\bibinfo{author}{M.~J. Barber}, \bibinfo{author}{J.~W. Clark},
  \bibinfo{author}{C.~H. Anderson},
\newblock \bibinfo{title}{Neural representation of probabilistic information},
\newblock \bibinfo{journal}{Neural Comp.} \bibinfo{volume}{15}
  (\bibinfo{year}{2003}{\natexlab{a}}) \bibinfo{pages}{1843--1864}.
\bibitem[{Barber et~al.(2003{\natexlab{b}})Barber, Clark, and
  Anderson}]{BarClaAnd:2003a}
\bibinfo{author}{M.~J. Barber}, \bibinfo{author}{J.~W. Clark},
  \bibinfo{author}{C.~H. Anderson},
\newblock \bibinfo{title}{Generating neural circuits that implement
  probabilistic reasoning},
\newblock \bibinfo{journal}{Physical Review E (Statistical, Nonlinear, and Soft
  Matter Physics)} \bibinfo{volume}{68} (\bibinfo{year}{2003}{\natexlab{b}})
  \bibinfo{pages}{041912}.
\bibitem[{Eliasmith and Anderson(1999)}]{EliAnd:1999}
\bibinfo{author}{C.~Eliasmith}, \bibinfo{author}{C.~H. Anderson},
\newblock \bibinfo{title}{Developing and applying a toolkit from a general
  neurocomputational framework},
\newblock \bibinfo{journal}{Neurocomputing} \bibinfo{volume}{26}
  (\bibinfo{year}{1999}) \bibinfo{pages}{1013--1018}.
\bibitem[{Eliasmith and Anderson(2002)}]{EliAnd:2002}
\bibinfo{author}{C.~Eliasmith}, \bibinfo{author}{C.~H. Anderson},
  \bibinfo{title}{Neural Engineering: The Principles of Neurobiological
  Simulation}, \bibinfo{publisher}{{MIT} Press}, \bibinfo{year}{2002}.
\bibitem[{Barber et~al.(2003)Barber, Clark, and
  Anderson}]{barber/clark/anderson:2003a}
\bibinfo{author}{M.~J. Barber}, \bibinfo{author}{J.~W. Clark},
  \bibinfo{author}{C.~H. Anderson},
\newblock \bibinfo{title}{Neural representation of probabilisitic information},
\newblock \bibinfo{journal}{Neural Comp.} \bibinfo{volume}{15}
  (\bibinfo{year}{2003}) \bibinfo{pages}{1843--1864}.
\bibitem[{Papoulis(1991)}]{papoulis:1991}
\bibinfo{author}{A.~Papoulis}, \bibinfo{title}{Probability, Random Variables,
  and Stochastic Processes}, \bibinfo{publisher}{McGraw-Hill, Inc.},
  \bibinfo{address}{New York, NY}, \bibinfo{edition}{third} edition,
  \bibinfo{year}{1991}.

\end{thebibliography}







\end{document}